\documentstyle[prl,aps]{revtex}

\begin{document}
\draft

\twocolumn[\hsize\textwidth\columnwidth\hsize\csname  @twocolumnfalse\endcsname 
\vskip2pc]

{\bf D'Anna {\it et al}. Reply}{\em :} In the preceding Comment\cite{AoCom},
P. Ao affirms that his vortex many-body theory\cite{AoVarious} explains the
observations reported in a recent Letter\cite{DAnnaHallPRL}. Indeed, we
provide here some arguments to show the lack of a satisfactory explanation
within the known theoretical frameworks.

We start by summarizing our results. For the magnetic field tilted away from
the c-axis more than about 2${{}^{\circ }}$, with decreasing temperature we
observe, first, the Hall sign reversal to negative values just below $T_{c}$%
, and, secondly, a sharp decrease toward large negative values of the Hall
conductivity at the vortex-lattice melting temperature $T_{m}$. The Hall
angle tends to small values (Fig. 1 in the Letter).

We have two main scenarios to explain our data. One follows the vortex {\it %
many-body} idea\cite{AoVarious} in which vortex-lattice defects provide the
negative Hall contribution. The other neglects vortex-vortex interactions%
\cite{variousSingleVortex} and, in consequence, explains the Hall behavior
in terms of {\it microscopic} electronic processes which affect the
vortex-core. Both explanations lead to difficulties, or need speculative
arguments to fit the data.

The difficulty with the many-body scenario is that the negative Hall
contribution arises from vortex-lattice defects. For the defects to exist,
the vortex-lattice must exist, at least locally, and must be pinned. But the
negative Hall contribution starts already above $T_{c}$ since the Hall
voltage starts to deviate from the positive normal state level above $T_{c}$%
, as shown for example in Fig. 1. In consequence, the many-body scenario
requires a locally ordered vortex-lattice and pinning at $T>T_{c}$ which is
difficult to accept. Nevertheless, one can speculate that vortex-lattice
defects become relevant for the Hall (and longitudinal) behavior {\it below}
the vortex-lattice melting transition. Just below the melting transition
vortex-lattice defects might contribute to the anelastic or plastic motion
under the effect of large dc currents, as the noisy Hall angle and Hall
conductivity observed in Fig. 1 seem to suggest, but this scenario is not
considered in ref.\cite{AoVarious} and it requires further development.

In the single-vortex scenario the Hall conductivity is closely related to
microscopic electronic processes which determine the vortex core structure.
In the BCS dirty limit scenario, and assuming a particle-hole asymmetry such
that the vortices are negatively charged, the hydrodynamic contribution can
drive the vortex Hall conductivity to a sign opposite to the normal state%
\cite{variousSingleVortex}, possibly accounting for the Hall sign reversal,
but severe discrepancies with the doping dependence remain unexplained\cite
{NagaokaDopingPRL}.

An alternative microscopic explanation can be constructed considering
negatively charged preformed electron-pairs which Bose condense at $T_{c}$%
\cite{GIL97}. In a pure Bose-Einstein condensation scenario, the core
contributions to the Hall force and damping coefficients are absent, and one
expects a large zero-temperature vortex Hall conductivity and Hall angle.
This Bose-Einstein scenario is tempting because it is consistent with our
finding of a very large Hall conductivity in the vortex-solid. But the
immediate consequence will be that the vortex-lattice melting transition has
a {\it microscopic origin,} that is the Bose condensation of preformed
bosons. Although this supports the similar conclusion deduced from the
observation that critical amplitude fluctuations persist almost down to the
vortex-lattice melting transition\cite{CooperPRL}, it remains speculative
and further theoretical work is necessary. Moreover, the Hall angle is
small, suggesting that the core damping term is relevant, possibly because
of the $d$-wave nature of the vortex-core in high-$T_{c}$ superconductors.

In conclusion, we think that the contested affirmation that no existing
theoretical model fully explains our observations is true. We recognize that
the vortex many-body theory may explain part of our data, notably the
contribution of vortex-lattice defects to the Hall conductivity in the solid
phase, but fails to account for the liquid and fluctuation region behavior.
On the other hand, a microscopic single-vortex theory taking into account
the $d$-wave pairing symmetry and details of the Fermi surface might explain
the Hall behavior in the liquid and solid phase. But at this stage, these
are speculative scenarios.

\bigskip 

G. D'Anna,$^{1}$ V. Berseth,$^{1}$ L. Forr\'{o},$^{1}$ A. Erb,$^{2}$ and E.
Walker$^{2}$

$^{1}$IGA, DP, Ecole Polytechnique F\'{e}d\'{e}rale de Lausanne, CH-1015
Lausanne, Switzerland

$^{2}$DPMC, Universit\'{e} de Gen\`{e}ve, CH-1211 Gen\`{e}ve, Switzerland

\smallskip

PACS numbers: 74.25.Fy, 74.60.Ge, 74.72.Bk

\end{document}